\documentclass[%
pra%
,reprint%
,twocolumn%
,secnumarabic%
,floatfix%
,superscriptaddress, showpacs%
,amssymb, amsmath, nobibnotes, aps]{revtex4-1}

\usepackage[latin1]{inputenc}
\usepackage{epsfig}
\usepackage{color}
\usepackage{ulem}

\newcommand{\ind}[1]{_{\text{#1}}}
\newcommand{\bra}[1]{\left<#1\right|}
\newcommand{\ket}[1]{\left|#1\right>}

\begin{document}

\title{Semi-analytical model for nonlinear light propagation in strongly interacting Rydberg gases}
\author{Martin G\"arttner}
\affiliation{Max-Planck-Institut f\"{u}r Kernphysik, Saupfercheckweg 1, 69117 Heidelberg, Germany}
\author{Shannon Whitlock}
\affiliation{Physikalisches Institut, Universit\"at Heidelberg, Im Neuenheimerfeld 226, 69120 Heidelberg, Germany}
\author{David W.\ Sch\"{o}nleber}
\author{J\"org Evers}
\affiliation{Max-Planck-Institut f\"{u}r Kernphysik, Saupfercheckweg 1, 69117 Heidelberg, Germany}



\date{\today}

\begin{abstract}
Rate equation models are extensively used to describe the many-body states of laser driven atomic gases. We show that the properties of the rate equation model used to describe nonlinear optical effects arising in interacting Rydberg gases can be understood by considering the excitation of individual super-atoms. From this we deduce a simple semi-analytic model that accurately describes the Rydberg density and optical susceptibility for different dimensionalities. We identify the previously reported universal dependence of the susceptibility on the Rydberg excited fraction as an intrinsic property of the rate equation model that is rooted in one-body properties. Benchmarking against exact master equation calculations, we identify regimes in which the semi-analytic model is particularly reliable. The performance of the model improves in the presence of dephasing which destroys higher order atomic coherences.
\end{abstract}

\maketitle

\section{Introduction}

Atoms in highly excited states, Rydberg atoms, exhibit strong and long-range interactions combined with long lifetimes. Due to strong interaction induced level shifts, driving these atoms with coherent laser fields (for example under electromagnetically induced transparency (EIT) conditions) gives rise to strong optical nonlinearities~\cite{schempp2010, pritchard2010}. These nonlinearities can be exploited to engineer non-classical states of light \cite{peyronel2012, dudin2012b, maxwell2013}, to produce entangled states of single photons and atomic excitations \cite{li2013}, to introduce interactions between photons in an atomic medium \cite{firstenberg2013}, or to observe sub-Poissonian statistics of dark-state polaritons \cite{hofmann2012}.
 
The theoretical modeling of ensembles of long-range interacting particles coherently interacting with light fields far from equilibrium turns out to be quite challenging. Mean field models fail quickly as the atomic density increases, and due to the exponentially growing Hilbert space, exact treatments are limited to small atom numbers \cite{robicheaux2005, younge2009, gaerttner2012}. In the EIT configuration, where the Rydberg level is accessed via a resonantly coupled intermediate level, incoherent processes such as the spontaneous decay of the intermediate level are crucial and make a full master equation (ME) treatment necessary. 

Various approaches have been pursued to tackle this problem \cite{reslen2011, gorshkov2011,gorshkov2012, pritchard2012, sevincli2011,sevincli2011b, petrosyan2011, petrosyan2012, ates2011,heeg2012,schoenleber2013}, one of them being the so called rate equation (RE) approach \cite{ates2011, heeg2012,ates2007a, ates2007b,  hoenig2013, gaerttner2013, schempp2013,schoenleber2013}. This approach may be used to calculate the steady state of a cloud of two- or three-level atoms subject to near resonant laser driving by adiabatic elimination of all coherences and including interactions as level shifts of the Rydberg level of individual atoms. This classical model is fast, simple to implement and allows to treat up to $10^5$ atoms and is thus directly applicable to many experimental setups. It has proven to perform well by comparison to experimental data \cite{hofmann2012, gaerttner2013, schempp2013} and predicts many remarkable properties such as the universal relation between the nonlinear optical susceptibility and the Rydberg excited fraction in three-level media. One question that immediately arises is, whether these properties are only valid if the system behaves classically due to strong decoherence or if they also hold in largely coherent systems. In other words: In which parameter regimes is the RE model valid?

In this paper we investigate the application of the RE model to an ensemble of resonantly driven three-level atoms. Through extensive Monte-Carlo simulations, Ates {\it et al.}~\cite{ates2011, sevincli2011b} found that the nonlinear optical susceptibility and the Rydberg density obey a universal relation. We show that this universal relation can be understood as an intrinsic property of the RE model. Then we show that the global observables, predicted by the RE model can be reproduced using a simple semi-analytical model by treating the atomic medium as a collection of super-atoms~\cite{parigi2012, stanojevic2013}. Starting from the analytic solution of the RE for a single super-atom, we illustrate how the RE model is capable of reproducing the scaling laws predicted in~\cite{weimer2008}.
Combining the universal relation~\cite{ates2011} with our result for the Rydberg excited fraction, we can describe the nonlinear optical response of a resonantly driven Rydberg gas in terms of single atom properties alone. To check the validity of the semi-analytical model we benchmark it by comparing to full ME simulations with few atoms. We find that it performs well whenever interatomic coherences are suppressed.

\section{Rate equation model}
\label{sec:re_model}

The RE model provides a way to calculate the steady state of a strongly interacting many-body system subject to laser driving, with a computational complexity that scales almost linear with the atom number as long as the Rydberg excited fraction is small. For details of the algorithm, see Refs.~\cite{ates2007a, ates2011, heeg2012}. The setup we have in mind is the three-level system, for example in $^{87}$Rb, with atomic states $|g\rangle,|e\rangle,|R\rangle$ as used in Ref.~\cite{hofmann2012}. The ground state $\ket{g}=\ket{5S_{1/2}}$ is resonantly coupled to an intermediate state $\ket{e}=\ket{5P_{3/2}}$ by a (weak) probe laser with Rabi frequency $\Omega_p$. The intermediate state $\ket{e}$ is resonantly coupled to the Rydberg state $\ket{R}=\ket{55S_{1/2}}$ by a (strong) coupling laser with Rabi frequency $\Omega_c$. The $\ket{e}$ state spontaneously decays to the ground state with rate $\Gamma$, while the Rydberg state is long lived (decay rate much smaller than $\Gamma$). The additional dephasings caused by the finite laser bandwidths lead to the total linewidths $\gamma_{eg}$ and $\gamma_{gR}$ of the probe transition and the two photon transition, respectively~\cite{hofmann2012}. Two atoms that are in the Rydberg state experience repulsive van der Waals type interactions with strength $C_6=50\,$GHz~$\mu$m$^6$.

The Hamiltonian of an ensemble of $N$ atoms, in the rotating wave approximation, reads ($\hbar=1$)    
\begin{equation}
H=\sum_{i=1}^N H_L^{(i)} + \sum_{i<j}\frac{C_6 \ket{R_i R_j}\bra{R_i R_j}}{|\mathbf{r}_i-\mathbf{r}_j|^6} 
 \label{eq:Hamiltonian}
\end{equation}
where $H_L^{(i)}=\Omega_p/2 \ket{g_i}\bra{e_i} + \Omega_c/2 \ket{e_i}\bra{R_i} + h.c.$ describes the coupling of the atoms to the laser fields. In this work we restrict our analysis to the case of resonant laser excitation, however, the RE approach is also straightforwardly applied to the case of off-resonant driving \cite{gaerttner2013, schempp2013}.
Incoherent processes can be included as Lindblad terms \cite{fleischhauer2005} leading to the master equation for the density matrix
\begin{equation}
\label{eq:masterequation}
 \dot{\rho}=-i[H,\rho]+\mathcal{L}[\rho].
\end{equation}
For a single atom ($N=1$) one can transform this into a RE for the populations of the atomic levels by adiabatically eliminating the coherences ($\dot{\rho}_{ab}=0$ for $a\neq b$) \cite{ates2007a}. In the many-body case one can generalize this to a RE for the populations of the many-body configurations $\boldsymbol{\sigma} = \{\sigma_1,\sigma_2,\ldots ,\sigma_N\}$, where $\sigma_i\in\{g,e,R\}$ \cite{heeg2012}. We employ a Monte Carlo technique to solve the many-body RE. That is, starting from the global ground state $\{g,g,\ldots ,g\}$ we perform a random walk through the space of configurations $\boldsymbol{\sigma}$ and average over many such trajectories, ensuring the convergence to a global steady state. The Hamiltonian $H$ and the Lindblad term $\mathcal{L}[\rho]$ couple two configurations $\boldsymbol{\sigma}$ only if they differ solely in the state of one atom. Therefore, in order to calculate the transition rates between two states, one only has to take into account the jump probability of the atom whose state is changing. The interaction between atoms in the Rydberg state is incorporated as a shift of the Rydberg level of the considered atom, $\Delta\ind{int}^{(i)}=\sum_{j\neq i}^{\prime} C_6/|r_{i}-r_{j}|^6$, where the primed sum only runs over atoms that are in the Rydberg state in the current step. $\Delta\ind{int}^{(i)}$ enters into the single atom master equation as a detuning of the Rydberg level of atom $i$, (i.e., via an additional term $H_\Delta^{(i)}=\Delta\ind{int}^{(i)}\ket{R_i}\bra{R_i}$), thereby excluding true many-body quantum correlations.

Technically, in a single step of the Monte Carlo procedure, an atom is picked randomly, its transition rates to other states are calculated by solving the single atom master equation including the local interaction shift, and in a further random process the decision is made on whether the atom changes its state and, if it does, which state it jumps to. 
We use the steady-state values of the populations of the considered atom as jump probabilities as done in Ref.~\cite{heeg2012}, but different from Refs.~\cite{ates2007a, ates2011, schempp2013}. This is justified as long as the solution is restricted to the global steady state. This enables us to simulate atomic ensembles consisting of up to $5\times10^4$ atoms in reasonable computation times at the cost of abandoning any physical dynamics.

\section{Origin of the universal relation}
\label{sec:univ_scal}

In this section we show how the universal relation reported in Ref.~ \cite{ates2011} arises as an intrinsic feature of the RE model.
This relation connects the steady state value of the imaginary part of the nonlinear optical susceptibility $\chi=\mathrm{Im}[\chi_{ge}]\propto \mathrm{Im}[\rho_{eg}]$ to the Rydberg density. It was found heuristically from many-body simulations of the RE model~\cite{ates2011} and reads
\begin{equation}
\label{eq:univ_scal_perfectEIT}
 \chi/\chi_{2L} = \frac{f_{bl}}{1+f_{bl}} \,.
\end{equation}
The blockade fraction is defined as
\begin{equation}
\label{eq:fbl}
 f_{bl}=\rho_{RR}^{(0)}/\rho_{RR}-1 \, ,
\end{equation}
where $\rho_{RR}$ is the steady-state Rydberg excited fraction (or Rydberg density, if multiplied by the atomic density $n_0$) and $\rho_{RR}^{(0)}=\rho_{RR}(\Delta\ind{int}=0)$ is its value in the non-interacting limit. $\chi_{2L}$ is the value of $\chi$ in the absence of the coupling laser ($\Omega_c=0$), which corresponds to the case of a two-level atom with states $\ket{g}$ and $\ket{e}$ only. In the non-interacting limit (reached for low atomic densities and/or probe intensities), the Rydberg density is equal to the non-interacting Rydberg density ($f_{bl}\rightarrow 0$). Hence $\chi=0$, corresponding to perfect EIT. In the strongly-interacting limit, when the Rydberg population is strongly suppressed due to the Rydberg blockade ($\rho_{RR}\ll \rho_{RR}^{(0)}$), the absorption approaches its two-level value $\chi/\chi_{2L}=1$.

The RE model always fulfils the universal relation~\eqref{eq:univ_scal_perfectEIT} due to the following reasoning: In steady state the susceptibility $\chi$ is proportional to the population of the intermediate level $\rho_{ee}$ \cite{ates2011}. Substituting $\chi/\chi_{2L}=\rho_{ee}/\rho_{ee}^{2L}$ into Eq.~\eqref{eq:univ_scal_perfectEIT}, the relation is simply 
\begin{equation}
\label{eq:univ_scal_perfectEIT2}
 \rho_{ee}/\rho_{ee}^{2L}=1-\rho_{RR}/\rho_{RR}^{(0)} \,.
\end{equation}
If this linear relation between the steady-state population of the intermediate level and the Rydberg level holds for the single-atom steady states for any (interaction induced) detuning $\Delta\ind{int}$, then it also holds for the ensemble average obtained from the RE model due to the fact that in the RE model interactions are only taken into account as level shifts of the Rydberg level. Indeed for the case of perfect EIT ($\gamma_{gR}=0$) we find that both sides of Eq.~\eqref{eq:univ_scal_perfectEIT2} are equal to
\begin{equation}
\frac{4\gamma_{ge}\Delta^2\ind{int}(\Gamma\gamma_{ge}+2\Omega_p^2)}{4\gamma_{ge}\Delta^2\ind{int}(\Gamma\gamma_{ge}+2\Omega_p^2) + (\Omega_c^2+\Omega_p^2)(\Gamma\Omega_c^2+\gamma_{ge}\Omega_p^2)},
\end{equation}
in the single atom case. Thus the universal relation \eqref{eq:univ_scal_perfectEIT} is found to be an intrinsic property of the RE model that arises due to the fact that in the RE model interactions are included as shifts in the two-photon detuning.

\begin{figure}[t]
  \centering
 \includegraphics[width=\columnwidth]{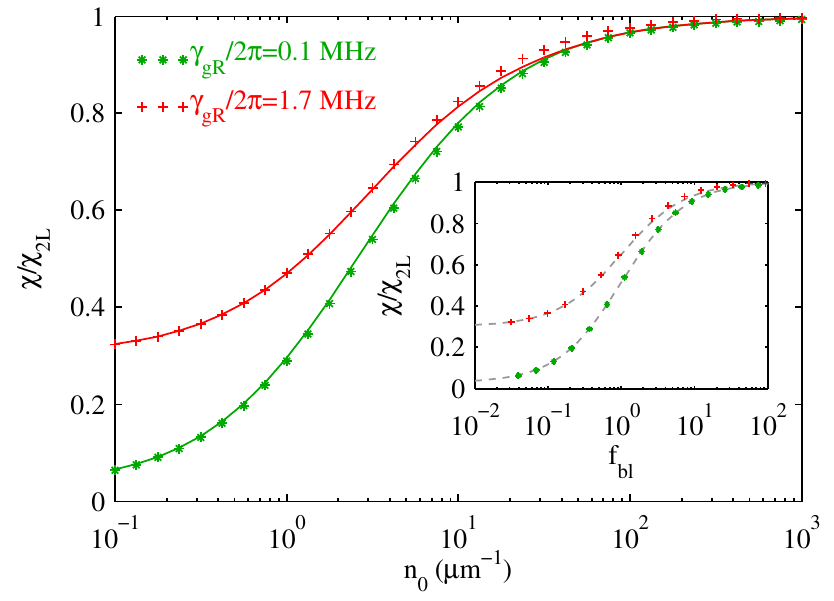}
 \caption{Rescaled absorption for varying atomic density. The symbols show data from RE simulations while the solid lines are the result of a semi-analytical model described in Sec.~\ref{sec:semianalyticmodel}. Parameters as in \cite{hofmann2012}: $C_6=50$\,GHz$\mu$m$^6$, $\Omega_p=1\,$MHz, $\Omega_c=5.1$\,MHz, $\Gamma=6.06$\,MHz, dephasing of the probe transition 0.1(0.33)\,MHz, dephasing of the two photon transition 0.1(1.7)\,MHz for the green (red) points. The inset shows the comparison between the RE model results and the modified universal relation~\eqref{eq:univ_scal}.}
 \label{fig:chi_universal}
\end{figure}
If one considers non-zero $\gamma_{gR}$ caused by laser dephasing or decay from the Rydberg level which suppress the $gR$-coherence, then $\chi$ will be non-zero even in the non-interacting limit (imperfect EIT).  To account for this, the universal relation can be heuristically modified to
\begin{equation}
 \chi/\chi_{2L} = \frac{\chi_0/\chi_{2L}+f_{bl}}{1+f_{bl}} \,,
 \label{eq:univ_scal}
\end{equation}
where we denoted the residual susceptibility in the non-interacting limit ($\Delta\ind{int}=0$) with $\chi_0$. This result can also be obtained from a simple hard sphere picture \cite{sevincli2011b, parigi2012, tanasittikosol2011, stanojevic2013}.
For imperfect EIT there is no such analytic equality as the one shown above for the perfect EIT case. However, as long as dephasing and Rydberg state decay are small, the modified universal relation \eqref{eq:univ_scal} is fulfilled to a good approximation.

Figure \ref{fig:chi_universal} shows the result of RE simulations for a homogeneous (periodic boundaries) one-dimensional sample of $5\times10^4$ atoms. The density is varied by reducing the system length. The coupling parameters are similar to those used in Ref.~\cite{hofmann2012}. The inset shows the comparison between the results of the RE model and the modified universal relation~\eqref{eq:univ_scal}. We observe that for low dephasing (green) the RE result follows the universal curve extremely well, while for relatively large dephasing (red) small deviations can be seen. The main figure shows the same results as a function of the atomic density $n_0$. The solid lines are the result of the semi-analytical model introduced in the next section which was used to obtain $f_{bl}$ as a function of $n_0$.

\section{Emergence of collective effects}
\label{sec:REmodel}

\subsection{High density regime}

\begin{figure}[t]
  \centering
 \includegraphics[width=\columnwidth]{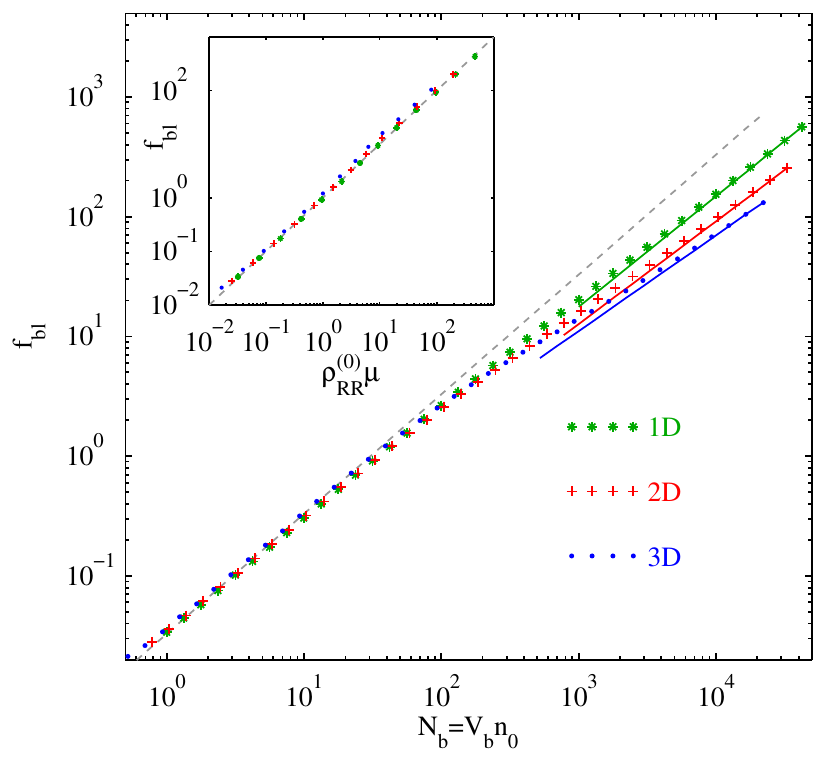}
 \caption{Blockade fraction as a function of atomic density for different trap geometries. The density is varied by changing the edge length of the cube/square/string for a fixed atom number of $N=5\times10^4$. The dashed line has unit slope $f_{bl}=\rho^{(0)}_{RR}N_b$. The solid lines are proportional to $N_b^\nu$ with $\nu=12/13$, $6/7$, and $4/5$, for 1D, 2D, and 3D, respectively. The inset shows $f_{bl}$ plotted as a function of $\rho^{(0)}_{RR}\mu$, where $\mu$ is the average super-atom size including collective enhancement of the atom-light coupling (Sec.~\ref{sec:coll_rb}).}
 \label{fig:fbl_var_dim}
\end{figure}

Having studied the connection between $\chi$ and $f_{bl}$ predicted by the RE model, we now focus on the dependence of $f_{bl}$ on the system parameters, in particular on the atomic density.
Figure~\ref{fig:fbl_var_dim} shows the dependence of $f_{bl}$ on the atomic density $n_0$ obtained from the RE model for systems of different dimensionality. The parameters are the same as for the red line in Fig.~\ref{fig:chi_universal} and are comparable to those of Ref.~\cite{hofmann2012} except for a higher probe Rabi frequency. We rescaled the atomic density by multiplying with the volume of a blockade sphere to give the number of atoms per blockade sphere, $N_b=n_0 V_b$. $V_b=\{2r_b,\pi r_b^2,4\pi r_b^3/3\}$ for \{1D, 2D, 3D\} respectively. The blockade radius $r_b$ is defined as the distance at which the two-body interaction energy $C_6/r^6$ equals the single-atom EIT resonance width $w_0=\gamma_{gR}+\Omega_c^2/\gamma_{eg}$ given by the single-atom steady-state population resonance $\rho_{RR}(\Delta\ind{int})$,
\begin{equation}
\label{eq:RbeqBinary}
 r_b=\left(\frac{2C_6\gamma_{eg}}{\gamma_{eg}\gamma_{gR}+\Omega_c^2}\right)^{1/6} \,.
\end{equation}
For the parameters of the simulation $r_b\approx 5\,\mu\mathrm{m}$.
Remarkably, $f_{bl}$ has a very simple dependence on $N_b$. In the low density regime all curves collapse to $f_{bl}=N_b  \rho^{(0)}_{RR}$ (dashed line), while in the high density limit $f_{bl}$ is proportional to $N_b^{2p/(2p+d)}$ (solid lines), where $d$ is the system dimensionality and $p$ is the exponent of the interaction potential. For a van der Waals potential one has $p=6$, yielding scaling exponents $12/13$, $6/7$, and $4/5$ for 1D, 2D, and 3D, respectively. It is somewhat surprising that these scalings, also discussed by Weimer {\it et al.} \cite{weimer2008} are reproduced by the RE model since they are rooted in the $\sqrt{N_b}$-enhancement of the Rabi-frequency due to coherent collective excitation that does not enter explicitly in the RE model.

\subsection{Semi-analytic model}
\label{sec:semianalyticmodel}

To understand how collective effects manifest themselves in the RE model, we regard the atomic ensemble as consisting of a collection of blockade spheres or super-atoms. An isolated super-atom has a certain steady-state probability $P_{ss}$ to be excited. The value of $P_{ss}$ predicted by the rate equation model can be calculated analytically. By equating the width of the super-atom resonance $P_{ss}(\Delta\ind{int})$ to the mean-field shift produced by surrounding super-atoms, we can self-consistently determine the size of the super-atoms. The resulting model closely reproduces the RE result for $f_{bl}$ over the full density range.

\subsubsection{Super-atom excitation probability}

To obtain the Rydberg excited fraction for a single super-atom comprising $N$ atoms only two configurations need to be distinguished: either all atoms are in noninteracting states ($\ket{g}$ or $\ket{e}$), or one of the atoms is in the Rydberg state. The second configuration is referred to as the excited state in the following. Multiple excitations are avoided due to the Rydberg blockade. We denote the probability to be in the excited state after $k$ Monte-Carlo steps by $P_{k}$. We now determine the probability $P_{k+1}$ to be in the excited state after the next Monte Carlo step. If the super-atom is in the excited state, the probability to de-excite it is $(1-\rho^{(0)}_{RR})/N$, where $\rho^{(0)}_{RR}$ is the single-atom steady-state population of the Rydberg level. Thus the probability to stay in the excited state is $1-(1-\rho^{(0)}_{RR})/N$. If no atom is excited after $k$ steps, the probability to excite one in the next step is $\rho^{(0)}_{RR}$. Putting this together the RE model follows the recursive relation
\begin{equation}
\label{eq:series}
 P_{k+1}=P_k\left( \frac{N-1+\rho^{(0)}_{RR}}{N} \right)+(1-P_k)\rho^{(0)}_{RR} 
\end{equation}
with $P_0=0$ assuming all atoms start in the ground state. For $k$ approaching infinity this series converges to the steady-state excitation probability 
\begin{equation}
\label{eq:PS}
 P_{ss}=\frac{\rho^{(0)}_{RR} N}{(N-1)\rho^{(0)}_{RR}+1}\,.
\end{equation}
Note that this result coincides with what is found by adiabatically eliminating all but the collective excited state and the ground state of a super-atom in the weak probe limit \cite{petrosyan2011,petrosyan2012b}, which explains why the RE model coincides with the super-atom model proposed in Ref.~\cite{petrosyan2011} (see also Ref.~\cite{gaerttner2013}). This expression is also consistent with $P_{ss}=N/(N+1)$ found in \cite{honer2011} for $\rho^{(0)}_{RR}=1/2$ (resonant driving and strong dephasing), and for the case of perfect EIT (no decay of the $gR$-coherence) it coincides with the steady-state population obtained for a single atom with $\sqrt{N}$-enhanced probe Rabi frequency.

\subsubsection{Size of the super-atoms in an extended sample}
\label{sec:coll_rb}

Equation~\eqref{eq:PS} gives the steady-state population of the Rydberg state for a given $N$. We now consider an extended atomic sample that is divided into super-atoms, each containing $N_c$ atoms. We denote the collectively enhanced blockade radius with $r_c$ and the mean number of atoms per collective blockade sphere $\langle N_c\rangle=4\pi r_c^3/3$ (for 3D). Due to local density fluctuations in an ensemble of randomly distributed atoms, $N_c$ will fluctuate from super-atom to super-atom. To account for this we perform an average over a Poissonian distribution
\begin{equation}
 \label{eq:Pavg}
 \langle P_{ss}\rangle = \sum_{N_c=0}^\infty{\frac{\mathrm{e}^{-\mu}\mu^{N_c}}{N_c!}P_{ss}}\,,
\end{equation}
where $\mu=\langle N_c\rangle$.
We note that for high atomic densities ($\mu\gg1$), or in the weak probe limit (where $\rho^{(0)}_{RR}\ll1$) the averaging does not have a significant influence (i.e. $\langle P_{ss}\rangle=P_{ss}$).

Finally, we determine the size of each super-atom (mean number of atoms per super-atom, $\mu$) in a self-consistent manner. For this we generalize Eq.~\eqref{eq:PS} by substituting $\rho^{(0)}_{RR}$ with $\rho_{RR}(\Delta\ind{int})$, i.e., by including a mean-field shift caused by the surrounding super-atoms. For single atoms the blockade radius is determined by equating the interaction shift to the resonance width $w_0$. Analogously, one equates the half width of the mean super-atom excitation probability $\langle P_{ss}(\Delta\ind{int})\rangle$ to the mean-field shift and solves for $\mu$ (or equivalently for $r_c$):
\begin{equation}
\label{eq:Rbeq}
 \langle P_{ss}(\Delta\ind{int}=w)\rangle=\frac{1}{2}\langle P_{ss}(\Delta\ind{int}=0)\rangle \,,
\end{equation}
where the interaction shift $w=C_6/r_c^6=w_0 (n_0 V_b)^2/\mu^2$ (for 3D), where we have used $w_0=C_6/r_b^6$. 

Numerically solving Eq.~\eqref{eq:Rbeq} for $\mu$, one obtains the size of the super-atoms as a function of $n_0$. Using $f_{bl}=\rho^{(0)}_{RR} \mu$ the dependence of $f_{bl}$ on $n_0$ is reproduced correctly over the full range of densities. This is illustrated in the inset of Fig.~\ref{fig:fbl_var_dim} where we plot the same data as in the main panel as a function of $\rho^{(0)}_{RR} \mu$. Indeed, all data points lie on the diagonal line over the full range of densities including the collective regime. We note a possible slight deviation at high densities (especially for the 3D geometry) where the RE results lie just above the diagonal, which we attribute to the breakdown of the super-atom picture due to the influence of overlapping blockade spheres. 

\subsubsection{Scaling laws}

Solving Eq.~\eqref{eq:Rbeq} in the limit $\mu\rightarrow\infty$ we find the following scaling behavior of collective blockade radius, Rydberg density, excited fraction, and blockade fraction ($f_{bl}\propto \rho_{RR}^{-1}$ for $f_{bl}\gg1$): 
\begin{subequations}
 \begin{align}
  r_c &\propto n_0^{-1/(2p+d)}\,, \\
  n_R \propto r_c^{-d} &\propto n_0^{d/(2p+d)}\,, \\
  \rho_{RR} = n_R/n_0 &\propto n_0^{-2p/(2p+d)} \,,\\
  f_{bl} \propto \rho_{RR}^{-1} &\propto n_0^{2p/(2p+d)}\,.
 \end{align}
\end{subequations}
This matches the scaling observed in Fig.~\ref{fig:fbl_var_dim} for $p=6$ (van der Waals) and $d=1,2,3$.

\section{Nonlinear light propagation}

The result of the preceding section can be exploited to describe the propagation of a resonant probe beam through a Rydberg EIT-medium. Combining the semi-analytical model of Sec.~\ref{sec:semianalyticmodel} with the universal relation of Eq.~\eqref{eq:univ_scal}, we obtain an expression for the local susceptibility $\chi(\mathbf{r})$ which only depends on the laser parameters, the interaction strength and the local atomic density $n_0(\mathbf{r})$. This is how the solid lines in the main panel of Fig.~\ref{fig:chi_universal} are obtained, showing nearly perfect agreement with the RE simulations. In Rydberg EIT, the probe beam is weak and becomes attenuated by absorption in the atomic medium. Under the assumption that the atomic steady state is reached and $n_0$ and $\Omega_p$ vary slowly in space, one can calculate the probe beam absorption by integrating
\begin{equation}
 \partial_z\Omega_p(\mathbf{r})=-\frac{k}{2}\chi[\Omega_p(\mathbf{r}),n_0(\mathbf{r})]\Omega_p(\mathbf{r})
 \label{eq:laser_attenuation}
\end{equation}
along the propagation direction $z$. For the setup of Ref.~\cite{hofmann2012} this model yields results identical to the RE simulations including the probe beam attenuation \cite{hofmann2012, gaerttner2013}. Compared to the RE model including probe beam attenuation~\cite{gaerttner2013}, the semi-analytic model allows to numerically calculate the probe beam absorption very efficiently.

\section{Comparison with exact master equation simulations}
\label{sec:benchmark}

In the following we test the validity of the semi-analytic model by comparing to full master equation simulations of the steady state of few-atom systems. By elaborating the characteristic features for the case of a fully blockaded sample we find that the model performs well in the presence of dephasing of the coherence $\rho_{gR}$ and in particular for asymmetric couplings ($\Omega_p\ll\Omega_c$ or $\Omega_p\gg\Omega_c$). In the limit of large dephasings, the classical rate equation model becomes exact~\cite{lesanovsky2013}. In the intermediate regime, however, the steady state Rydberg and intermediate state populations can differ between the two descriptions.

\begin{figure}[t]
  \centering
 \includegraphics[width=\columnwidth]{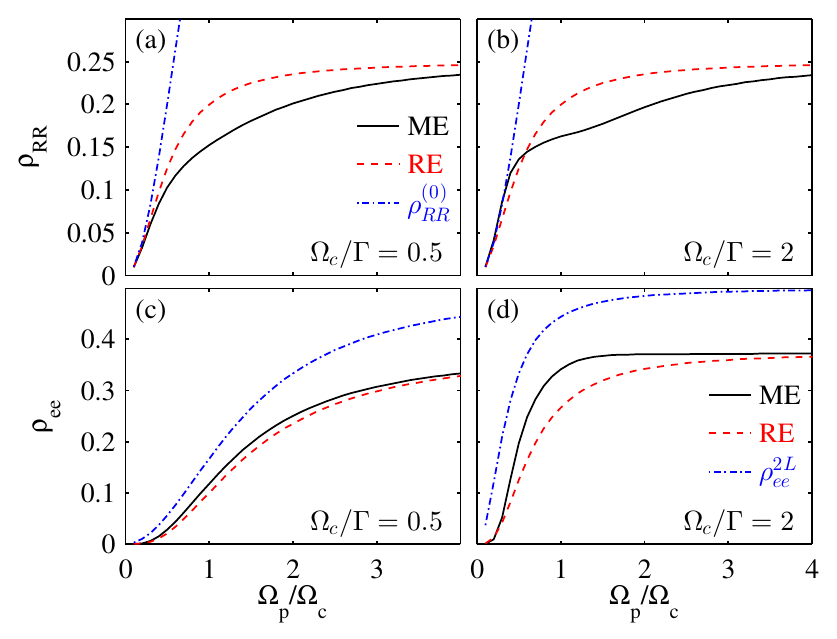}
 \caption{Steady-state Rydberg population $\rho_{RR}$ and intermediate state population $\rho_{ee}$ for an ensemble of $4$ perfectly blockaded atoms as a function of the probe Rabi frequency $\Omega_p$. Laser dephasings and decay from the Rydberg level are set to zero (perfect EIT). In (a) and (c), the coherent laser driving is weak compared to the decay rate $\Gamma$ of the intermediate level, in (b) and (d) the coherent driving is rather strong. Solid black lines show the exact ME solution, red dashed lines show the semi-analytic result. The blue dot-dashed lines show the non-interacting value of $\rho_{RR}$ and the two-level value (without Rydberg level) of $\rho_{ee}$, respectively.}
 \label{fig:MERE_pop}
\end{figure}

Figure~\ref{fig:MERE_pop} shows the dependence of Rydberg ($\rho_{RR}$) and intermediate state ($\rho_{ee}$) populations on the probe Rabi frequency ranging from the weak probe to the strong probe regime. We assume a fully blockaded sample of $N=4$ atoms under perfect EIT conditions (zero laser dephasings and no spontaneous decay from the Rydberg state). In the fully blockaded case with fixed $N$, the semi-analytical model is equivalent to the RE model and is given by Eq.~\eqref{eq:PS}. Figures.~\ref{fig:MERE_pop}(a) and (c) correspond to the regime of weak driving ($\Omega_c<\Gamma$). We observe that the Rydberg population predicted by Eq.~\eqref{eq:PS} (red dashed line) agrees well with the exact solution in both the weak and strong probe regimes, while for intermediate $\Omega_p/\Omega_c$ the Rydberg population is overestimated by the semi-analytical model. For strong coherent driving [Figs.~\ref{fig:MERE_pop}(b) and (d)] a similar situation is encountered, with the exception that for weak probe a region arises where the semi-analytical model underestimates the exact result. The intermediate state population [shown in Figs.~\ref{fig:MERE_pop}(c) and (d)] is underestimated in the intermediate regime.
We note that in the presence of laser dephasing all the deviations become smaller, but the qualitative features remain the same, as will be shown below.

\begin{figure}[t]
  \centering
 \includegraphics[width=\columnwidth]{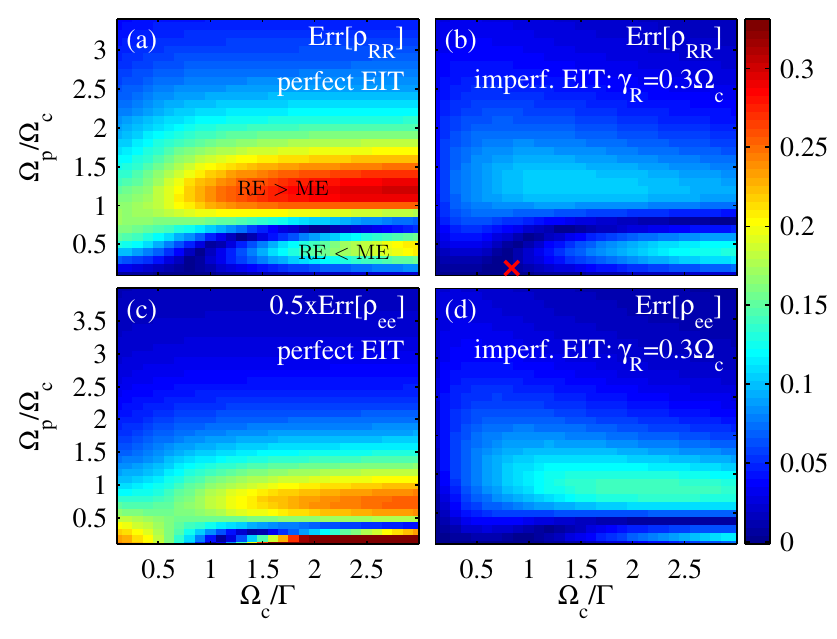}
 \caption{Absolute value of the relative deviation in Rydberg and intermediate state population between RE and ME for a fully blockaded ensemble of $N=2$ atoms. In each figure, the parameters are varied between weak and strong coherent drive from left to right, and between weak and strong probe from bottom to top. We use perfect EIT conditions for the left column, (a) and (c), while in the right column, (b) and (d), a dephasing of the coherences with the Rydberg level $\gamma_{R}=0.3\Omega_c$ is included. The red cross marks (approximately) the parameters used in Fig.~\ref{fig:fbl_var_dim}.}
 \label{fig:Err_all}
\end{figure}

Figure \ref{fig:Err_all} shows a systematic study of the relative deviation between the semi-analytical model and the ME simulations for the case of 2 atoms, which shows all qualitative features. Figures~\ref{fig:Err_all}(a) and (b) show the relative deviation in $\rho_{RR}$, while (c) and (d) show the deviation in $\rho_{ee}$, calculated from 
\begin{equation}
 \mathrm{Err}[\rho_{aa}] = \left|\frac{\rho_{aa}^{\mathrm{ME}}-\rho_{aa}^{\mathrm{RE}}}{\rho_{aa}^{\mathrm{ME}}}\right| 
\end{equation}
with $a\in\{R,e \}$, respectively. 
In all plots, two regions can be distinguished: In the weak probe regime, there is a region where the semi-analytical model underestimates the exact $\rho_{RR}$, and in the intermediate probe regime, the exact result is overestimated (and vice versa for $\rho_{ee}$). The two regions are separated by a line of vanishing deviation. The region where the Rydberg population is underestimated only exists for sufficiently strong coherent driving ($\Omega_c>\Gamma$). Comparing Figs.~\ref{fig:Err_all}(a) and (b) we observe that including decoherence effects such as finite laser linewidths, the deviations decrease, however the qualitative features persist. In (b) we included a dephasing of the Rydberg level with rate $\gamma_R=0.3\Omega_c$, which approximately corresponds to the parameters used in Sec.~\ref{sec:REmodel}. The cross marks the parameters used in Fig.~\ref{fig:fbl_var_dim}, showing that for these parameters the semi-analytical model performs very well. The deviations in $\rho_{ee}$ look quite similar, however, here now the population is overestimated by the semi-analytical model for weak probe and underestimated for intermediate and strong probe intensity. Additionally, the weak probe feature is much narrower than for $\rho_{RR}$. The large deviations observed at weak probe and strong coherent driving are an artefact of the idealized case of perfect EIT. As soon as  a dephasing of the Rydberg state is present, the good performance of the semi-analytic model is recovered at $\Omega_p/\Omega_c\ll 1$, cf.\ Fig.~\ref{fig:Err_all}(d).
For larger atom numbers ($N=3,4$) the qualitative features remain the same. The main difference is a vertical compression of the structures which amounts to a rescaling of $\Omega_p$ by a factor of $\sqrt{N}$. This means that in order to estimate the validity of the semi-analytical model for higher atom numbers, one should consider the parameter $\sqrt{N}\Omega_p/\Omega_c$ rather than $\Omega_p/\Omega_c$.

\section{Summary}
\label{sec:summary}

In summary we have shown that the RE model is a powerful tool for modelling the many-body physics of long range interacting three-level atoms in the frozen gas regime as long as inter-atomic coherences are suppressed. The smallness of these coherences can be due to strong dephasing effects, or due to strongly differing sizes of the Rabi frequencies, i.e., strong probe ($\Omega_p\gg\Omega_c$) or weak probe ($\Omega_p\ll\Omega_c$). We have found that the RE model reproduces collective effects that can be understood in terms of classical statistical arguments and resembles the expected steady state for a strongly dissipative fully blockaded ensemble. At low atomic densities the predictions of the RE model are consistent with a linear dependence of $f_{bl}$ on the density of ground state atoms. At high densities, i.e., in the collective regime, the RE model reproduces the scaling exponents that emerge due to collective effects. Generalizing the simple relation for $f_{bl}=\rho_{RR}^{(0)}N_b$ by including the collective blockade radius instead of the single-atom one, we can reproduce the results of the RE model over the full range of atomic densities.

The universal relation between the electromagnetic susceptibility $\chi$ and the blockade fraction $f_{bl}$ has been shown to be an intrinsic feature of the RE model that can be deduced from the single atom master equation and appears in RE simulations due to the fact that interactions are included only as level shifts. It was shown that this feature together with the relation $f_{bl}=\rho_{RR}^{(0)}\mu$ can be exploited to design a simple model for light propagation through Rydberg EIT media. 

Despite the fact that collective scaling exponents are reproduced by the RE and semi-analytic models, it is not immediately clear that the results obtained in this regime are quantitatively consistent with fully correlated many-body calculations. In order to check this, we have carried out full master equation calculations with few atoms for comparison. The semi-analytical model performs well in the presence of dephasing and in particular in the strong probe and very weak probe regimes, i.e., when $\rho^{(0)}_{RR}$ is either close to one or close to zero. In these cases the inter-atomic coherences are small and thus fragile with respect to dephasing.

\begin{acknowledgments}
We thank 
M.\ Fleischhauer, 
G.\ G\"unter, 
K.\ P.\ Heeg, 
C.\ S.\ Hofmann, 
M.\ H\"oning, 
M. Robert-de-Saint-Vincent, 
H.\ Schempp, 
and 
M.\ Weidem\"uller
for discussions. This work was supported by University of Heidelberg (Center for Quantum Dynamics, LGFG).
\end{acknowledgments}

\end{document}